\documentclass[prb,twocolumn,superscriptaddress]{revtex4}
\usepackage[T1]{fontenc}
\usepackage{color}
\usepackage{graphicx}
\usepackage{amsmath}
\usepackage{amssymb}
\usepackage{amsfonts}
\usepackage[colorlinks,hyperindex]{hyperref}
\usepackage{epsfig}

\def\beq{\begin{equation}}
\def\eeq{\end{equation}}

\begin{document}

\title{Critical dynamics and tree-like spatiotemporal patterns in exciton-polariton condensates}

\author{Nataliya Bobrovska}
\affiliation{Institute of Physics Polish Academy of Sciences, Al. Lotnik\'ow 32/46, 02-668 Warsaw, Poland}

\author{Andrzej Opala}
\affiliation{Institute of Physics Polish Academy of Sciences, Al. Lotnik\'ow 32/46, 02-668 Warsaw, Poland}

\author{Pawe{\l} Mi\k{e}tki}
\affiliation{Institute of Physics Polish Academy of Sciences, Al. Lotnik\'ow 32/46, 02-668 Warsaw, Poland}

\author{Micha{\l} Kulczykowski}
\affiliation{Institute of Physics Polish Academy of Sciences, Al. Lotnik\'ow 32/46, 02-668 Warsaw, Poland}

\author{Piotr Szymczak}
\affiliation{Institute of Theoretical Physics, Faculty of Physics, University of Warsaw, Pasteura 5, 02-093 Warsaw, Poland}

\author{Michiel Wouters}
\affiliation{TQC, Universiteit Antwerpen, Universiteitsplein 1, B-2610 Antwerpen, Belgium}

\author{Micha{\l} Matuszewski}
\affiliation{Institute of Physics Polish Academy of Sciences, Al. Lotnik\'ow 32/46, 02-668 Warsaw, Poland}

\begin{abstract}
  We study nonresonantly pumped exciton-polariton system in the vicinity of the dynamical instability threshold. We find that the system exhibits unique and rich dynamics, which leads to spatiotemporal pattern formation. The patterns have a tree-like structure, and are reminiscent of structures that appear in a variety of soft matter systems. Within the approximation of slow and fast time scales, we show that the polariton model exhibits self-replication point in analogy to reaction-diffusion systems. 
\end{abstract}

\maketitle
\section{Introduction} \label{sec:introduction}

Semiconductor exciton-polaritons are quantum quasiparticles that exist in structures where strong light-matter coupling overcomes decoherence~\cite{Kavokin_Microcavities}. Properties of microcavity polaritons, which combine the extremely low effective mass of confined photons with strong interactions of excitons, makes them an ideal candidate for studying quantum fluids of light~\cite{Carusotto_QuantumFluids}. Rapid progress in studies of these systems has led to observations of remarkable phenomena, such as nonequilibrium Bose-Einstein condensation~\cite{Deng_Condensation,Kasprzak_BEC}, quantum vortices~\cite{Deveaud_Vortices,Vina_VorticesCoherently,Sanvitto_InteractionsScatteringVortices,Yamamoto_VortexPair}, superfluidity~\cite{Amo_Superfluidity}, and Berezinskii-Kosterlitz-Thouless phase transition~\cite{Szymanska_NonequilibriumBKT,Sanvitto_TopologicalOrder}.

Several recent experiments provided evidence of dynamical instability in exciton-polariton condensates in the case of nonresonant pumping~\cite{Bobrovska_DynamicalInstability,Estrecho_SingleShotCondensation,Bloch_UnstableRegimes}. This instability is an inherent property of the open-dissipative Gross Pitaevskii model, widely used for describing the dynamics of these systems~\cite{Wouters_Excitations}. Signatures of instability were observed both in the case of organic microcavities~\cite{Bobrovska_DynamicalInstability}, as well as inorganic GaAs microcavities pumped continuously~\cite{Bloch_UnstableRegimes} and with ultrashort pulses~\cite{Estrecho_SingleShotCondensation}.

Despite these experimental observations, most studies of polariton fluids to date have focused on the stable regime of condensation. In particular, properties of the system close to instability threshold have not been a topic of a detailed study. This is of practical importance, since both stable and unstable regimes of condensation have been demonstrated experimentally~\cite{Bloch_UnstableRegimes,Estrecho_SingleShotCondensation,Sanvitto_TopologicalOrder,Bobrovska_DynamicalInstability}. It was pointed out that this regime can be characterized by interesting chaotic dynamics with unusual momentum distribution of fluctuations~\cite{Bobrovska_Adiabatic}. Note that chaotic evolution has been recently predicted to occur also in a polariton model with resonant plane wave driving~\cite{Gavrilov_SpinTurbulence,Gavrilov_Chimeras}.

Spatial pattern formation in polariton systems has been studied in a number of different configurations both theoretically and experimentally~\cite{Baumberg_PhaseTransitions,Baumberg_GeometricallyLockedVortexLattices,Berloff_VortexLattices,Berloff_PatternFormation,Egorov_PatternsMotion,Manni_PatternFormation,Sanvitto_XWaves,Lederer_PolaritonPatterns,Kwong_OpticalSwitchingPatterns,Marchetti_SpontaneousPatternsCoherently,Krizhanovskii_PatternFormationStatistical,Bramati_CoherentMerging,Bramati_AnnularVortexChain,Bramati_InteractionShapedLattices,Ostrovskaya_TalbotEffect}. In this work, we investigate dynamical behavior close to the instability threshold in detail, and predict {\it spatiotemporal} pattern formation.
We find that the dynamics results in tree-like structures in space-time coordinates, which exhibit branching, or self-replication. The behavior of the system becomes very similar to that occurring in certain soft matter systems, including combustion~\cite{Moses_FingeringInstability}, bacterial growth~\cite{Golding_BacterialGrowth}, chemical reactions~\cite{Lee_PatternsChemicalFronts}, wetting films~\cite{Safran_FingeringThinFilms}, or self-replicating pattern formation in general diffusion-reaction models~\cite{Reynolds_SelfReplicatingPatterns}. We describe the physical mechanism responsible for branching, resulting from phase separation into condensed and uncondensed regions. In analogy to reaction-diffusion systems, the existence of two time scales, corresponding to slow evolution and fast splitting dynamics, allows to understand the occurrence of self-replication and determine the threshold for its occurence.
As a result, we find that polariton systems in the critical regime display rich dynamics that is very different from superfluid behavior observed in the stable regime. We also provide an analogy to extensively studied soft-matter systems.

We discuss the experimental observation of splitting dynamics. We point out that while direct detection of branching would be difficult in experiment due to the chaotic nature of the process and the picosecond time scales involved, it is possible to observe signatures of branching in second-order spatiotemporal correlation function. This method allows to perform time-averaged experiment in which many branching events occurring in a condensate over a long acquisition time contribute to a nontrivial pattern of spatiotemporal correlations, which can be considered a smoking gun of branching dynamics.

\section{Model} \label{sec:model}
We model evolution of an exciton-polariton condensate   
using the open-dissipative Gross-Pitaevskii equation (ODGPE) for the wavefunction $\psi$, coupled to 
the rate equation for the density of exciton reservoir, $n_R$. In our work, we will focus mainly on the one dimensional case, when the condensate is trapped in a 1D microwire~\cite{Bloch_ExtendedCondensates,Bloch_UnstableRegimes}. Results in the two dimensional case are briefly discussed in Appendix~\ref{sec:2D}. The 1D evolution equations read
\cite{Wouters_Excitations}
\begin{equation}
\begin{split}
\label{GP}
i \hbar\frac{\partial \psi}{\partial t} &=-\frac{\hbar^2 D}{2 m^*} \frac{\partial^2 \psi}{\partial x^2} 
+ g_{\rm C} |\psi|^2 \psi + g_{\rm R} n_{\rm R} \psi \\
&+i\frac{\hbar}{2}\left(R n_{\rm R} - \gamma_{\rm C} \right) \psi , \\
\frac{\partial n_{\rm R}}{\partial t} &= P(x) - (\gamma_{\rm R}+ R |\psi|^2) n_{\rm R}, 
\end{split}
\end{equation}
 where $P(x)$ is the exciton creation rate determined by the pumping profile, $m^*$ 
is the effective mass of lower polaritons, $D=1-iA$ where $A$ is a dimensionless constant accounting for kinetic energy relaxation,
$\gamma_{\rm C}=\tau_C^{-1}$ and $\gamma_{\rm R}=\tau_R^{-1}$ are the polariton and exciton loss rates related to their lifetimes $\tau_{C,R}$, 
and $(R,g_i)=(R^{\rm 2D},g_i^{\rm 2D})/\sqrt{2\pi d^2}$ are
the rates of stimulated scattering into the condensate and the interaction coefficients, 
rescaled in the one-dimensional case~\cite{Bobrovska_Stability}, where $d$ is of the order of the microwire width.

In a model without noise, a nonzero homogeneous stationary solution of the above model can be found in the form $\psi(x,t)=\psi_0e^{-i\mu_0t}$, $n_R(x,t)=n_R^0$.
This solution exists above threshold pumping $P>P_{\rm th}=\gamma_C\gamma_R/R$ and is given by $|\psi_0|^2=(P/\gamma_C)-(\gamma_R/R)$, $n_R^0=\gamma_C/R$, and $\mu_0=g_C|\psi_0|^2+g_Rn_R^0$.
This homogeneous solution becomes dynamically unstable (via Benjamin-Feir instability) in a certain parameter range, as 
predicted~\cite{Wouters_Excitations,Ostrovskaya_DarkSolitons,Bobrovska_Stability} and recently observed experimentally~\cite{Bobrovska_DynamicalInstability,Estrecho_SingleShotCondensation,Bloch_UnstableRegimes}. 
The criterion for linear stability in the case $A=0$ was derived in~\cite{Ostrovskaya_DarkSolitons,Liew_InstabilityInduced}
\begin{equation}
\label{condition}
\frac{P}{P_{\rm th}}>\frac{g_R}{g_C}\frac{\gamma_C}{\gamma_R}.
\end{equation}
In the case when $A\neq0$ linear stability can be determined by solving Bogoliubov eigenvalue problem numerically.

We note that the dynamics predicted in this paper appear to be quite general, and not limited to the model described above. As we demonstrate in Appendix A, the physics described occurs as well in a model of polaritons in a semimagnetic microcavity. This model does not include a reservoir explicitly, and the second degree of freedom is provided by the magnetization of magnetic ions.

\begin{figure*}
	\begin{center}
	\includegraphics[width=1.45\columnwidth]{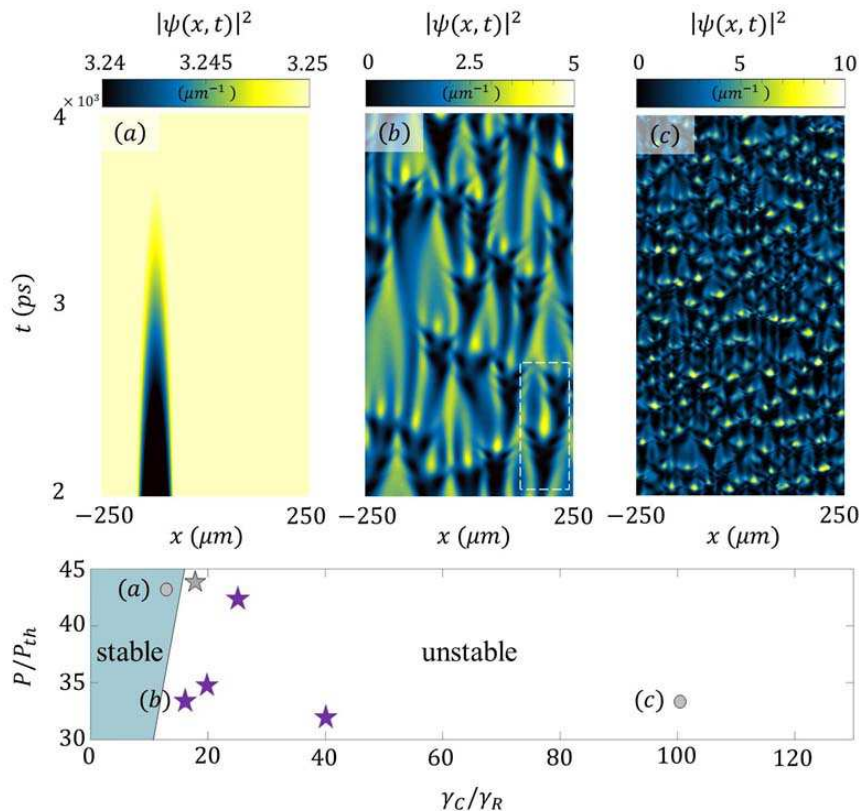} 
		\caption{Spatiotemporal pattern formation. Upper panels show evolution of condensate density in (a) the stable regime, (b) the critical regime, close to instability threshold, and (c) the deep unstable regime, as indicated in the stability diagram below. Panel (b) reveals spontaneous formation of the tree-like spatiotemporal patterns.
		The grey star in the phase diagram below corresponds to the case where phase turbulence is observed~\cite{Aranson_CGLEWorld}. Purple stars correspond to cases with clear tree-like branching evolution. 
		Parameters are $m^*=3.5\times10^{-5}m_e^0$, $\tau_R=1000$ps,  $\tau_C^{(a)}=76.92$ps, $\tau_C^{(b)}=62.01$ps, $\tau_C^{(c)}=9.95$ps, 
		$d=4\mu m $, 
		$g_C^{1D}=1.08$$\mu$eV$\mu m^2$,
		$g_R^{1D}=4g_C^{1D}$, 
		$R^{1D}=4.3 \times 10^{-3} \frac{\mu m}{ps}$, $A=0.9$.}
		\label{fig1}
	\end{center}
\end{figure*}

\section{Results} \label{sec:numerical}

Figures~\ref{fig1}(a)-(c) present examples of numerical dynamics of the ODGPE model~(\ref{GP}) (a) in the stable regime, (b) in the critical-unstable regime close to the stability threshold of Eq.~(\ref{condition}), and (c) in the deep unstable regime. We assume a small white noise in the polariton and reservoir fields at $t=0$, and a constant homogeneous pumping $P>P_{\rm th}$ for $t>0$.  In Figure~\ref{fig1}(a) typical behavior expected for the stable regime is visible, where initial condensate density fluctuation decays over time. In Figure~\ref{fig1}(c), an apparently random pattern of high density peaks is formed, as could be also expected in the unstable case.  On the other hand, in the intermediate case Fig.~\ref{fig1}(b) the instability apparently leads to pattern formation and spatiotemporal chaos, which takes the form of tree-like branching of domains which are characterized by low condensate density. We verified that such patterns appear in a relatively wide region of parameter space in the vicinity of the critical threshold. However, the estimation of exact limits of this region is a nontrivial task which will be postponed for a future study. The tree-like patterns are reminiscent of those occurring in certain soft-matter systems~\cite{Moses_FingeringInstability,Golding_BacterialGrowth,Lee_PatternsChemicalFronts,Safran_FingeringThinFilms,Reynolds_SelfReplicatingPatterns}. Below we draw an analogy between diffusion-reaction systems and critical dynamics of the polariton model. We note that somewhat similar patterns were recently predicted to occur in a complex Ginzburg-Landau equation (CGLE) polariton model, incorporating a carefully engineered complex periodic potential~\cite{Savenko_Reconstruction}. However this regime appears to resemble spatiotemporal intermittency regime of the CGLE~\cite{VanHecke_Intermittency} rather than dynamics of diffusion-reaction systems. 

\begin{figure}
	\begin{center}
	\includegraphics[width=\columnwidth]{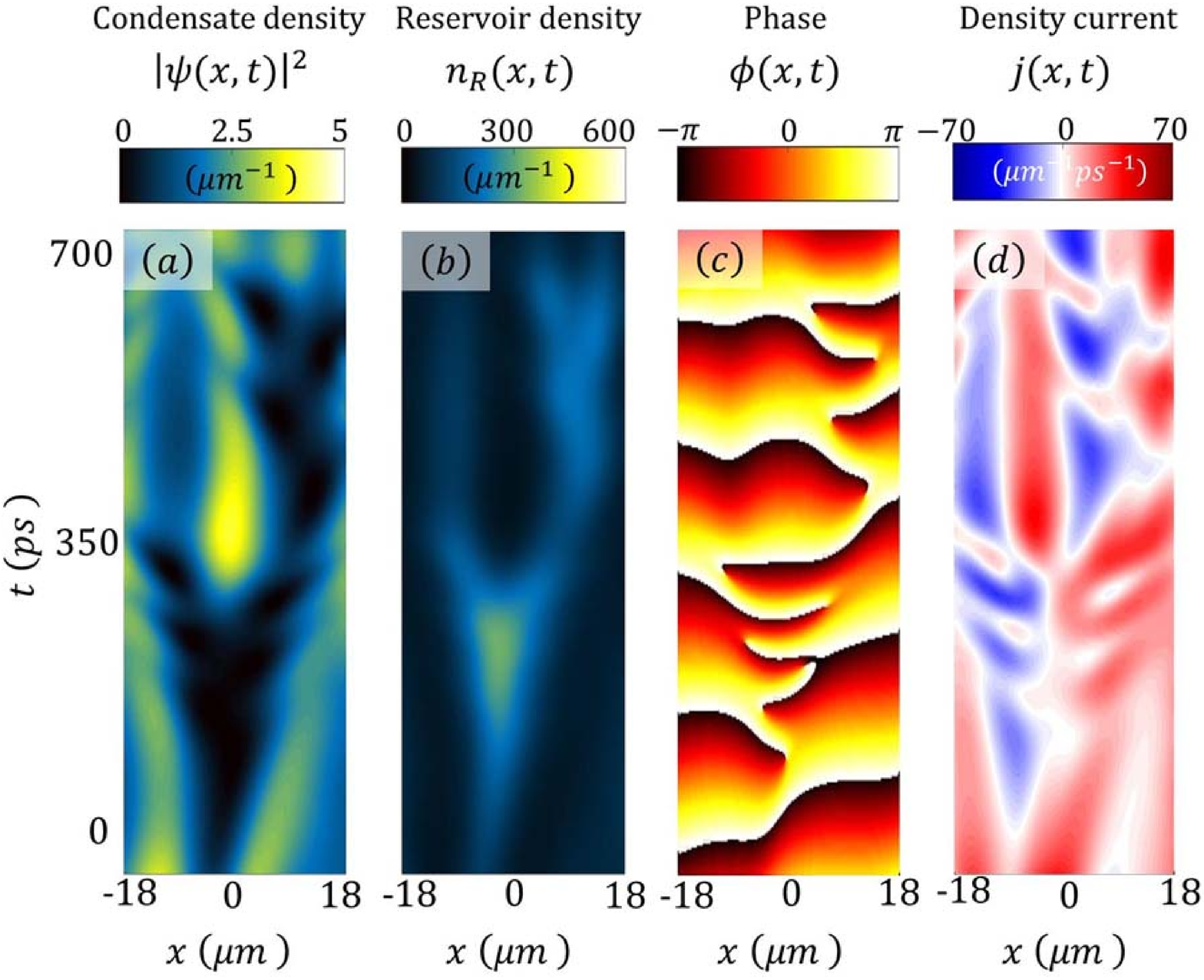} 
		\caption{Example of (a) condensate density, (b) reservoir density, (c)
		condensate wavefunction phase, and (d) condensate density current. The above figures correspond to a single branching event, selected from Fig.~\ref{fig1}(b) (marked with a white dashed box).}
		\label{fig2}
	\end{center}
\end{figure}

To investigate the dynamics of branching in more detail, we plot the evolution of condensate density and phase, together with the reservoir density for a single branching ``event'' in Fig.~\ref{fig2}. The correspondence between the regions of low condensate density and high reservoir density is a signature of phase separation, resulting from the repulsive polariton-reservoir interaction term $g_{\rm R}$ in Eqs.~(\ref{GP}). Phase separation is the driving force of dynamical instability in a polariton system~\cite{Bobrovska_Stability,Bobrovska_Adiabatic}. Here, it leads to the formation of well defined regions of high condensate density, separated from regions of high reservoir density, and the formation of separate branches visible in Fig.~\ref{fig2}. At the same time, it does not lead to a complete decay of the condensate into small lumps, as in the deep unstable regime of Fig.~\ref{fig1}(c), since rather wide regions of almost homogeneous condensate can still be distinguished between the branches. The existence of such two qualitatively different ``phases'' of low and high condensate density, corresponding to the branches and the regions between them, can be justified by the existence of two stationary homogeneous solutions of Eqs.~(\ref{GP})
\begin{align}
(a)\quad &|\psi|^2 = 0, &n_{\rm R}&=\frac{P}{\gamma_R}, \\
(b)\quad &|\psi|^2 = \frac{P}{\gamma_C} - \frac{\gamma_R}{R}, &n_{\rm R}&=\frac{\gamma_C}{R}, \nonumber
\end{align}
i.~e.~the zero solution and the nonzero stationary solution. While both these (spatially infinite) solutions are not stable in the unstable regime of condensation, the dynamics of the system appears to locally follow the form of either (a) or (b). This is confirmed by the magnitude of condensate and reservoir density in the branches and between them, which are close to values given by (a) and (b), respectively. 

The corresponding phase $\phi$ of the condensate wavefunction $\psi=|\psi|{\rm e}^{i\phi}$ is shown in Fig.~\ref{fig2}(c). Notice that the phase gradient in the time direction is different on the left and right hand side of the branch, as follows from the different frequency of $2\pi$ rotations of the phase along the time axis. This evidences the lack of phase coherence between the condensate regions on the two sides. In other words, the condensates which exist between the branches form uncorrelated condensate islands with no mutual phase coherence, but with coherence within each condensate. The branches, on the other hand, are regions where there is almost no condensate density and no phase coherence, which is visible as multiple phase discontinuities (spatiotemporal vortices) appearing in Fig.~\ref{fig2}(c).

\begin{figure}
	\begin{center}
	\includegraphics[width=\columnwidth]{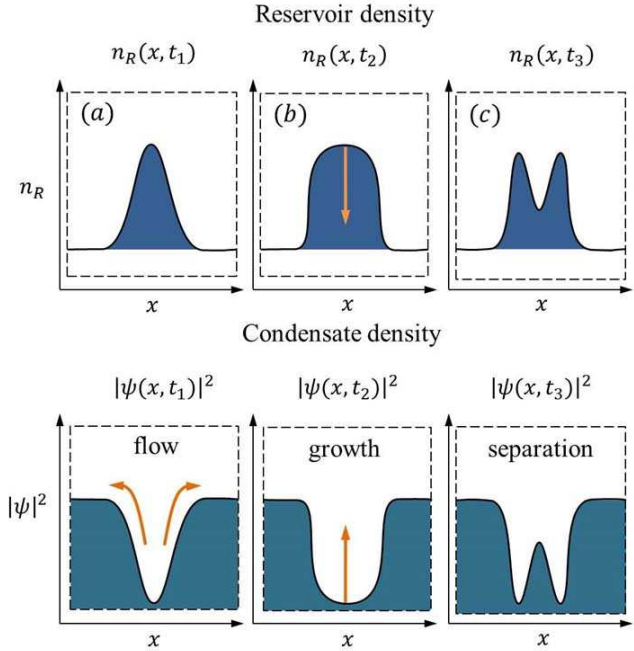} 
		\caption{Schematic illustration of the physical mechanism leading to branch splitting. (a) The repulsive potential generated by the reservoir-dominated branch leads to expulsion of polaritons from the branch, as well as growth of its dimensions. (b) When the branch becomes wide enough, the region in its center with a flat section of the potential becomes a seed for a new condensate island. (c) Condensate density quickly grows, leading to  separation of the two new branches.}
		\label{fig3}
	\end{center}
\end{figure}

The above observations, together with phase gradients in $x$ direction shown in Fig.~\ref{fig2}, allow for the understanding of physical mechanism of branching. Density current of polaritons can be calculated from the standard formula  $j=-i\hbar/2 m^* (\psi^*\partial \psi/\partial x - {\rm c.c.})$, and is plotted in Fig.~\ref{fig2}(d). A single branch before splitting is characterized by flux of polaritons from inside the branch to the outside regions, as shown schematically in Fig.~\ref{fig3} (left). This results from the repulsive potential $g_R n_R$ in Eq.~(\ref{GP}), created by the increased reservoir density in the (a) phase inside the branch. Indeed, above threshold $P>P_{\rm th}$ reservoir density is always higher in phase (a) than in phase (b). In the stable regime, this repulsive potential is screened by the lower condensate density, which acts through the condensate self-interaction term $g_C|\psi|^2$. However, as we enter the unstable regime, the reservoir-induced repulsive potential begins to dominate, and leads to outflow of condensate density from the regions of increased reservoir density, resulting in phase separation.

The outside directed flow of polaritons from inside the branch results in gradual increase of the spatial extent of the branch, as shown in the middle panels of Fig.~\ref{fig3}, which is also visible as widening of the branch in time in Fig.~\ref{fig2}(a). However, the spatial extent cannot increase indefinitely, since the (a) phase inside the branch is not a stable state. When the branch becomes wide enough, dynamical instability sets in, leading to splitting of the branch into two. The stability of the branch below a certain spatial extent of the branch and instability above this extent is a crucial property which makes the tree-like dynamics possible. When the branch splits, it develops a small high condensate density area in its center, which grows quickly thanks to the spontaneous scattering from the reservoir to the condensate. This is possible as the reservoir density is locally high, and the outflow of polaritons is suppressed locally thanks to the flattening of the effective potential as shown in Fig.~\ref{fig3} (middle). The fast growth of condensate density leads to the formation of two separate branches as depicted in Fig.~\ref{fig3} (right).

\begin{figure}
	\begin{center}
	\includegraphics[width=\columnwidth]{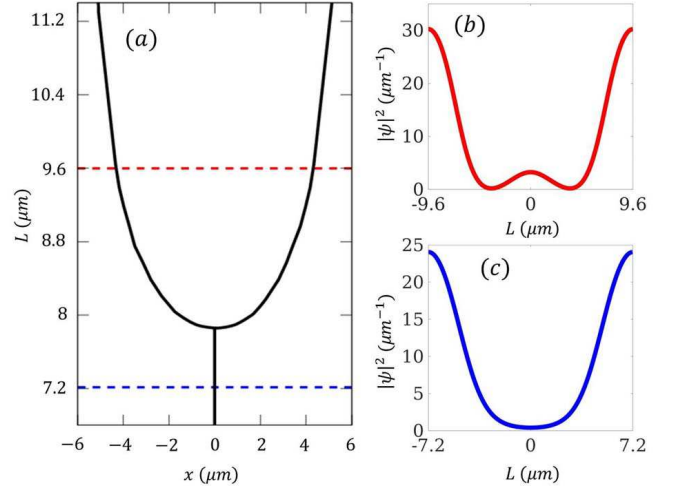} 
		\caption{Self-replication transition occurring when the the finite system size is increased. Due to the periodic boundary conditions, this corresponds to increasing the distance between neighboring branches. Solid line in panel (a) shows the position of a minimum (or two minima) of condensate density in a steady state, calculated in a box of size $L$. This corresponds to a chain of equally spaced branches separated by a distance $L$. Two representative states shown in panels (b) and (c) correspond to dashed lines in panel (a). The splitting occurs as the distance to neighboring branches becomes larger than $L_{\rm threshold}\approx7.8 \mu$m. Parameters are $g_C^{1D}=0.76$ $\mu$eV$\mu$m, $g_R^{1D}=2g_C^{1D}$, $\tau_C=4$ps, $\tau_R=3.42$ps, $R^{1D}=0.19$ $\frac{\mu m}{ps}$ d=2$\mu$m, $P/P_{th}=1.5$, $A=0.05$.}
		\label{fig4}
 	\end{center}
\end{figure}

\begin{figure*}
	\begin{center}
	\includegraphics[width=1.4\columnwidth]{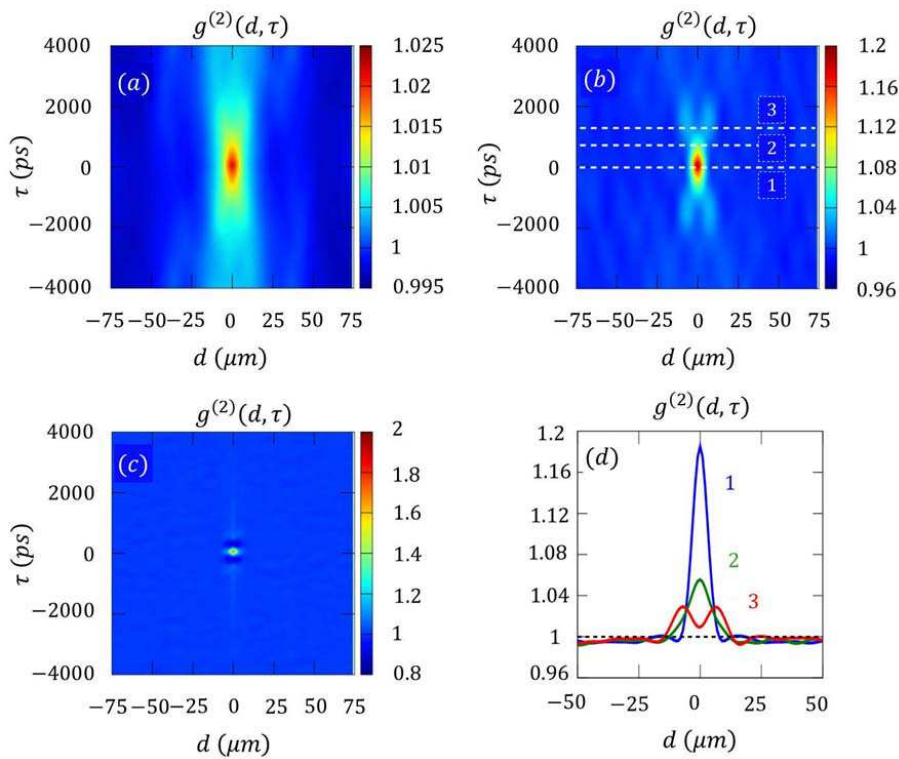} 
	\caption{Second order correlation function $g^{(2)}(d,\tau)$ is depicted for the stable (a), critical (b), and unstable (c) cases of Fig.~\ref{fig1}. The critical case (b) with branching density patterns is characterized by nontrivial spatiotemporal correlations which cannot be factorized into independent spatial and temporal parts. The characteristic ``horn'' features are signatures of branching in density evolution from Fig.~\ref{fig1}(b). In panel (d), cross-sections of $g^{(2)}(d,\tau)$ in the case (b) are plotted for $\tau=0$, $1000$ an $1500$ ps. Note that $g^{(2)}(0,0)$ is approximately equal to unity in (a) and equal to two in (c), which correspond to a coherent state and a classical random state (or thermal state), respectively.}
		\label{fig5}
	\end{center}
\end{figure*}

To describe the physics of splitting more quantitatively, we employ the time scale separation method, introduced in the study of  dynamics of self-replicating patterns in diffusion-reaction systems~\cite{Reynolds_SelfReplicatingPatterns}. This approach is based on the assumption that the evolution occurs on two different time scales. The slow movement of branches is occasionally interrupted by fast dynamics of splitting, or self-replication. Within this approach, the solutions in the slow phase of motion can be found approximately by assuming a steady state which consists of a chain of identical branches or a single branch within a finite box with periodic boundary conditions~\cite{Reynolds_SelfReplicatingPatterns}. The threshold of splitting can be determined from stability properties of this periodic solution. Such an approximation, although clearly not adequate to exactly describe the dynamics of non-periodic arrangement of branches as in Fig.~\ref{fig1}(b), allows to gain insight into the main mechanism driving the branching dynamics and determine the approximate threshold.

As obtaining an exact analytical solution is not viable in our nonlinear system, we employ numerical method based on the evolution of Eq.~(\ref{GP}) in a box of length $L$ with periodic boundary conditions. After sufficiently long time of evolution, we obtain a stationary stable solution. To minimize transient effects and avoid possible effects of multistability, we perform the simulations adiabatically, by feeding the result of one simulation as a starting point of another, with a slightly modified extent of the box $L$. This allows to follow one stable branch of solutions, and by changing $L$ in both directions we can detect the possible effects of bistability. We show the results of our investigation in Fig.~\ref{fig4}. In panel~(a) the range of investigated box sizes is shown on the vertical axis, with the solid line showing the positions of either a single minimum or two minima of the solution. The splitting of the minimum into two occurs when the box size is equal to about $L_{\rm threshold}\approx7.8 \mu$m. The examples of solutions with a single and two minima are shown in Figs.~\ref{fig4}(b,c). These solutions resemble closely the density profiles obtained previously in a large system. The obtained threshold size of a branch before self-replication $L_{\rm threshold}$ is also in good agreement with typical spatial scales on which branching occurs in full simulations. At the same time, we did not observe any region in which the two kinds of solutions shown in Figs.~\ref{fig4}(b,c) would be stable for the same $L$.

\section{Detection of branching via correlations}

Direct observation of branching shown in Fig.~\ref{fig1}(b) would be a challenging task due to the short (picosecond) time scale of the dynamics. Although streak cameras can be used to observe polariton dynamics on such time scales, they usually require averaging over many repeated realizations of the experiment or over a relatively long acquisition time. Such methods would not provide evidence of branching due to the chaotic character of the process, in which patterns are expected to vary rapidly and from shot to shot. We propose to circumvent this problem by measuring second-order spatiotemporal correlations instead of emission intensity. The reasoning behind such approach is that even if branching  occurs at random positions and times, we can still recover its characteristic features in integrated correlation functions, since all branching events will contribute to it in a similar way. Second-order correlation function is defined as
\begin{equation}
  g^{(2)}(d,\tau)=\frac{\int |\psi(x,t)|^2|\psi(x+d,t+\tau)|^2dxdt}{(\int |\psi(x,t)|^2 dxdt)^2}
\end{equation}
where the spatial integral is taken over the size of the system. Time integration starts from the instant when the system achieves a quasi-stationary distribution, in which there are strong fluctuations, but observables have reached a steady state in a statistical sense. In practice, such state is established after several hundred picoseconds of evolution, when average density saturates.

In Figure~\ref{fig5} we visualize correlation functions corresponding to the three cases from Fig.~\ref{fig1}. Clearly, stable, critical and deep unstable cases are characterized by qualitatively different correlation functions. The characteristic horn-like shape of $g^{(2)}$ in Fig.~\ref{fig5}(b) is an indication of branching occurring in Fig.~\ref{fig1}(b). Note that the horns are directed both in positive and negative time direction, since $g^{(2)}(d,\tau)$ as defined above is a time- and space-symmetric function in the limit of infinite integration time. It is important to note that only the critical case Fig.~\ref{fig5}(b) is characterized by nontrivial spatiotemporal correlations. In both  Fig.~\ref{fig5}(a) and (c)  correlations can be approximately factorized into spatial and temporal functions, i.~e.~$g^{(2)}(d,\tau)\approx g^{(2)}_x(d) g^{(2)}_t(\tau)$, while such factorization is not possible in the case of  Fig.~\ref{fig5}(b). This is clearly shown in Fig.~Fig.~\ref{fig5}(d), where cross-sections of correlation function at three different values of $\tau$ are shown. On the other hand, we note that in the unstable case of Fig.~\ref{fig5}(c) temporal correlations $g^{(2)}_t(\tau)$ also have a nontrivial (non-Gaussian) character.

\section{Conclusions}

In conclusion, we demonstrated that nonresonantly pumped exciton-polariton condensates at the threshold of instability possess unique and rich dynamics, reminiscent of self-replicating patterns encountered in many soft-matter systems. 
We believe that these results provide an interesting link between quantum coherent wave systems and soft matter diffusion-reaction systems, which may stimulate further interaction between these areas of physics.

\acknowledgments

We thank Marzena Szyma\'nska and Sebastian Diehl for fruitful discussions. Support from National Science Centre, Poland Grants 2015/17/B/ST3/02273 and 2016/22/E/ST3/00045 is acknowledged.

\appendix 

\section{Branching in diluted magnetic
semiconductor model}

We discuss the generality of the observed effects. 
We find that branching appears not only in the open-dissipative Gross-Pitaevskii model with a reservoir, but also in a model of semimagnetic exciton-polaritons in which the reservoir is absent. 
The role of the reservoir is played by the collective magnetization of manganese ions coupled to the condensate.

Recently, experimental investigations of semimagnetic microcavities (Cd$_{1-x}$Mn$_x$Te) were performed~\cite{Mirek_AngularDependence,Pacuski_MagneticFieldEffect} in which quantum wells are doped with magnetic ions.
In these cavities, phenomena such as giant Zeeman splitting and polariton lasing were observed~\cite{Pacuski_StrongCoupling,Pietka_MagneticFieldTuning}. 
Magnetization of a diluted magnetic semiconductor is given by the Brillouin function $B_{\rm J}$~\cite{Gaj_Brillouin}
\begin{gather}
	\langle M(x, t) \rangle= n_{\rm M} g_{\rm M} \mu_{\rm B} J~B_{\rm J}\left( \frac{g_{\rm M} \mu_{\rm B} J  B_{\rm eff} }{k_{\rm B} T} \right),  \label{eq:Mmean}
\end{gather}
where $n_{\rm M}$ is the magnetic ion concentration, $g_{\rm M}$ is their g-factor, $J$ is the manganese total angular momentum equal to $5/2$, $\mu_{\rm B}$ is the Bohr magneton, $k_{\rm B}$ is the Boltzmann's constant, $T$ is the temperature of manganese ions.
$B_{\rm eff}=(\lambda/2)|\psi|^2$ is the effective magnetic field resulting from the presence of exciton-polariton condensate~\cite{Kavokin_InterplaySuperfluidityDMS}, with the strength of the ion-polariton coupling is denoted with $\lambda$. 
\begin{figure}[h]
    \centering
    \includegraphics[width=0.4\textwidth]{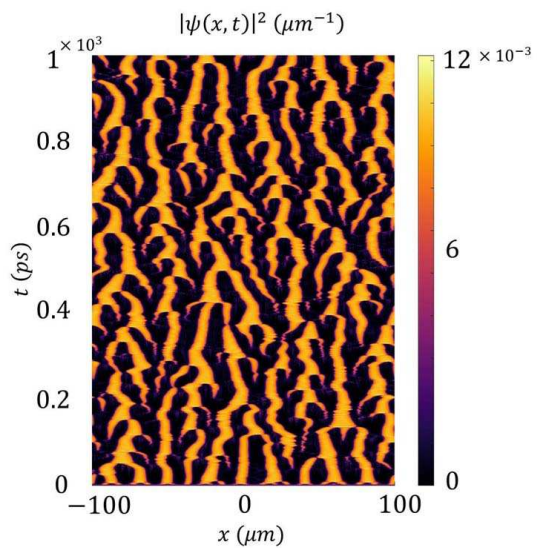}
    \caption{Example of branching evolution in the model of semimagnetic exciton-polaritons in which the second degree of freedom is due to the collective magnetization of manganese ions rather than the reservoir. Parameters are
    $\tau_M$ = $\mathrm{1.5 \times 10^{-14}}$ $\mathrm{s }$, 
    $g_{\rm 1}$ = $\mathrm{ 1.2 \times 10^{-9}}$ $\mathrm{meV~m}$, 
    $n_{\rm M}$ = $\mathrm{3.6 \times 10^{12}}$ $\mathrm{m^{-1}}$, 
    $B$ = $\mathrm{ 0~T }$, 
    $T$ = $\mathrm{ 0.1~K}$, 
    $P-\frac{1}{2}\gamma_{\rm L}$ = $\mathrm{ 6.6 \times 10^{-2}}$ $\mathrm{meV}$,
    $\gamma_{\rm NL}$ = $\mathrm{ 9.6 \times 10^{-12}}$ $\mathrm{meV~m}$, 
	$m^*$ = $\mathrm{10^{-5}}~ m_{\rm E}$,
	Rabi splitting $\Omega_R= 5$ meV. 
    }
    \label{fig:magnetic}
\end{figure}

Within the description in terms of complex Ginzburg Landau equation, which can be obtained from the full open-dissipative Gross-Pitaevskii model within the adiabatic approximation~\cite{Bobrovska_Adiabatic}, there is an additional effective potential caused by the ion-exciton interaction~\cite{Kavokin_InterplaySuperfluidityDMS, Mietki_MagneticPolaron}.
\begin{gather}
\begin{aligned}
i \hbar \frac{\partial \psi}{\partial t} &=  - \frac{\hbar^2}{2 m^*} \frac{\partial^2  \psi}{\partial x^2} + g_1 |\psi|^2 \psi  
+ iP\psi - \\ - &i\frac{1}{2}\gamma_{\rm L} \psi - i\gamma_{\rm NL} |\psi|^2 \psi - \lambda M \psi, \label{eq:CGLE1} 
\end{aligned}
\end{gather}
where the interaction between the polaritons with $g_1$, external pumping with $P$ and losses (linear and non-linear) with $\gamma_L$ and $\gamma_{NL}$.
We assume that circular pumping is homogeneous and the condensate remains circularly polarized, however the ion polarization is free to evolve.
Moreover, we introduce the spin relaxation time ($\tau_{\rm M}$) for magnetic ions. Then, the polariton evolution equation couples to the equation for manganese magnetization relaxation
\begin{gather}
\frac{\partial M(x,t)}{\partial t} = \frac{ \langle M(x, t) \rangle - M(x,t)}{\tau_{\rm M}} \label{eq:MTR1}
\end{gather}
We found that within this model, there exist a large region in parameter space in which tree-like branching occurs, and an example is shown in Fig.~\ref{fig:magnetic}.

\section{Two-dimensional case} \label{sec:2D}

\begin{figure}
    \centering
    \includegraphics[width=0.35\textwidth]{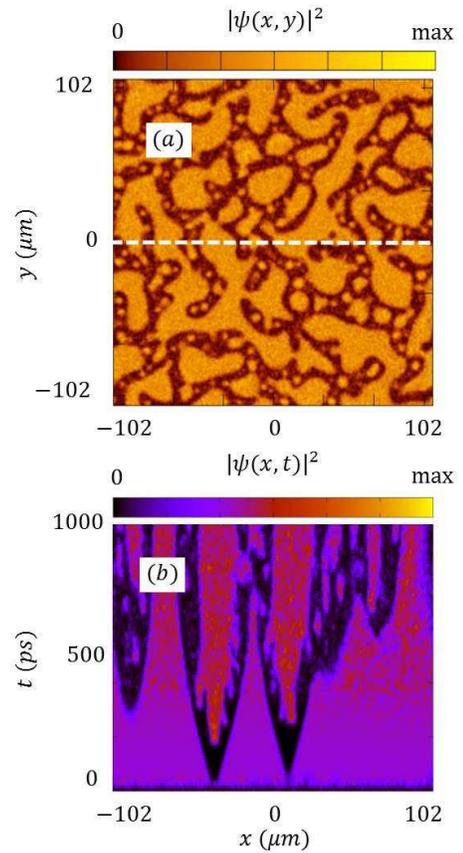}
    \caption{Example of branching evolution in the two-dimensional model. (a) Spatial pattern of the density of the condensate at $t_{\rm final}=10^3$ps. In panel (b), a cross-section of the density evolution at $y=0$ is shown, demonstrating branching patterns emerging from the initial state. Parameters in physical units are $m^*=5.5\times10^{-5}m_e^0$, $\tau_R=10\,$ps, $\tau_C=9\,$ps, $g_C=3.4$$\mu$eV$\mu$m$^2$, $g_R=6.8$$\mu$eV$\mu$m$^2$, $R=5.1$$\times 10^{-3}$ $\frac{\mu m^2}{ps}$, $L=204$$\mu$m, $A=0$.}
    \label{fig:2Df}
\end{figure}

To investigate whether dimensionality is an important factor in the occurrence of branching, we perform a series of numerical simulations in the two-dimensional extension of the model~(\ref{GP})
\begin{equation}
\begin{split}
\label{GP2D}
i \hbar\frac{\partial \psi}{\partial t} &=-\frac{\hbar^2 D}{2 m^*} \nabla^2 \psi
+ g_{\rm C} |\psi|^2 \psi + g_{\rm R} n_{\rm R} \psi +\\
&+i\frac{\hbar}{2}\left(R n_{\rm R} - \gamma_{\rm C} \right) \psi , \\
\frac{\partial n_{\rm R}}{\partial t} &= P(x) - (\gamma_{\rm R}+ R |\psi|^2) n_{\rm R}, 
\end{split}
\end{equation}
where $\nabla^2=\partial^2/\partial x^2+\partial^2/\partial y^2$, and initial conditions for the fields $\psi$ and $n_R$ are the same as before.
We found that for a similar range of parameters as in the 1D case, one can observe branching solutions as shown in Fig.~\ref{fig:2Df}. In panel (a), we show the density of the polariton condensate at a given time $t_{\rm final}$. The dynamics of branching is visible in panel (b), where a cross-section for $y=0$ is shown, demonstrating the formation of spatiotemporal patterns similar as in previous sections. The parameters of the simulation are given in the Figure caption. The lower quality of figures is due to the increased numerical mesh spacing in the 2D case, which was necessary because of the limited computational resources.

\clearpage

\bibliography{references}

\end{document}